\documentclass{conm-p-l}
\usepackage{amssymb,latexsym, amscd}
\usepackage[all]{xy}
\usepackage{graphicx}
 \newtheorem{theorem}{Theorem}[subsection]

 \newtheorem{proposition}[theorem]{Proposition}
 \theoremstyle{definition}
 \newtheorem{definition}[theorem]{Definition}
 \theoremstyle{definition}
 \newtheorem{example}[theorem]{Example}
 \theoremstyle{remark}
 
 \numberwithin{equation}{subsection}

\usepackage{latexsym}
\usepackage{amssymb}

\newcommand{\ben}{\begin{equation}}
\newcommand{\een}{\end{equation}}

\newcommand{\integer}{\ensuremath{{\mathbb Z}}}
\newcommand{\naturals}{\ensuremath{{\mathbb N}}}
\newcommand{\real}{\ensuremath{{\mathbb R}}}
\newcommand{\complex}{\ensuremath{{\mathbb C}}}

\newcommand{\rational}{\ensuremath{{\mathbb Q}}}

\newcommand{\U}[1]{\ensuremath{{\mathrm U( #1 )}}}

\newcommand{\GL}[1]{\ensuremath{{\mathrm {GL}_{ #1 }}}}
\newcommand{\GLC}[1]{\GL{#1}(\complex)}

\newcommand{\DD}{{\mathcal D}}
\newcommand{\PP}{{\mathcal P}}

\newcommand{\BB}{{\mathcal B}}

\newcommand{\UU}{{\mathcal U}}

\newcommand{\FF}{{\mathcal F}}
\newcommand{\GG}{{\mathcal G}}
\newcommand{\CC}{\mathcal{C}}
\newcommand{\II}{\mathcal{I}}
\newcommand{\LL}{\mathcal{L}}
\newcommand{\MM}{\mathcal{M}}
\newcommand{\HH}{\mathcal{H}}

\newcommand{\TT}{\mathcal{T}}
\newcommand{\PPP}{\mathcal{P}}

\newcommand{\Hom}{\mathrm{Hom}}
\newcommand{\Aut}{\mathrm{Aut}}

\newcommand{\Hol}{\mathrm{Hol}}

\newcommand{\Map}{\mathrm{Map}}

\newcommand{\Xx}{\mathsf{X}}

\newcommand{\Gg}{\mathsf{G}}
\newcommand{\Hh}{\mathsf{H}}
\newcommand{\Mm}{\mathsf{M}}
\newcommand{\Ss}{\mathsf{S}}
\newcommand{\Zz}{\mathsf{Z}}
\newcommand{\Yy}{\mathsf{Y}}

\newcommand{\Kk}{\mathsf{K}}

\newcommand{\target}{\mathsf{t}}
\newcommand{\source}{\mathsf{s}}

\newcommand{\Loop}{\mathsf{L}}
\newcommand{\Corr}{\ensuremath{{\mathbf{Corr}}}}

\newcommand{\virt}{\mathrm{virt}}

\newcommand{\Bun}{\mathrm{Bun}}

\newcommand{\To}{\longrightarrow}

\newcommand{\timests}{\: {}_{t}  \! \times_{s}}

\newcommand{\toparrow}[1]{\stackrel{#1}{\rightarrow}}

\newcommand{\ev}{\ensuremath{{\mathrm{ev}}}}

\newcommand{\Gr}[2]{\ensuremath{{\mathrm{Gr}_{#1}(#2)}}}

\newcommand{\M}{\mathrm{M}}

\newcommand{\obj}{\mathcal{O}bj }
\newcommand{\arr}{\mathcal{A}rr }

\newcommand{\ag}{\alpha}
\newcommand{\bg}{\beta}
\newcommand{\Sets}{\textbf{Sets}}
\newcommand{\Groups}{\textbf{Groups}}
\newcommand{\Abel}{\textbf{Ab}}
\newcommand{\Modules}{\textbf{Mod}}
\newcommand{\Rings}{\textbf{Ring}}
\newcommand{\Top}{\textbf{Top}}

\newcommand{\Cat}{\textbf{Cat}}
\newcommand{\Orbifolds}{\textbf{Orbifolds}}

\begin{document}

\title
{Topological Quantum Field Theories, Strings and Orbifolds.}

\author{Ernesto Lupercio and Bernardo Uribe}

\thanks{The first author was partially supported by CONACYT-M\'exico.}

\address{Departamento de Matem\'{a}ticas, CINVESTAV,
     Apartado Postal 14-740
     M\'{e}xico, D.F.  07000 M\'{E}XICO}
\email{lupercio@math.cinvestav.mx}
\address{Departamento de Matem\'{a}ticas, Universidad de los Andes,
Carrera 1 N. 18A - 10, Bogot\'a, COLOMBIA}
\email{buribe@uniandes.edu.co}

\begin{abstract}
In this  article, written primarily for physicists and geometers,
we introduce the notion of TQFT, orbifold, and then we survey the construction of TQFTs originating from orbifolds such as Chen-Ruan cohomology and orbifold string topology.
\end{abstract}

\dedicatory{Dedicated in memoriam to Prof. Raoul Bott.}

\maketitle

\tableofcontents

\section{Topological Quantum Field Theories}

In quantum mechanics one encounters the following situation.

\begin{example} Let $M$ be a Riemannian manifold (you  may want to think that it is $M=\real^3$).

 Given two points in $M$, say $p$ and $q$, we would like to compute the probability that a particle that starts in $p$ lands in $q$ after certain amount of time $T$. The answer is, of course, zero, but we can nevertheless still ask what is the probability that the particle will be at a distance less than $\epsilon$ from $q$.

Feynman gave a remarkable formula for the probability \cite{FeynmanQED}. Say that $\phi$ is the initial probability distribution for the position of the particle at $t=0$ (meaning that $\int_U |\phi_0|$ is the probability that the particle is in $U$ at $t=0$). Then the probability distribution $\phi_T$ for the position at $t=T$ is given by the \emph{path integral}
\begin{equation} \label{Feynman} \phi_T(q) = \int_{\PP_q} \phi_0(\gamma(0)) e^{-i \hbar S(\gamma)} \DD\gamma \end{equation}
where
$$ \PP_q = \{ \gamma \colon [0,T] \to M | \gamma(1) = q \} \subset \Map([0,T], M) $$
\begin{eqnarray*}
\includegraphics[height=2.0in]{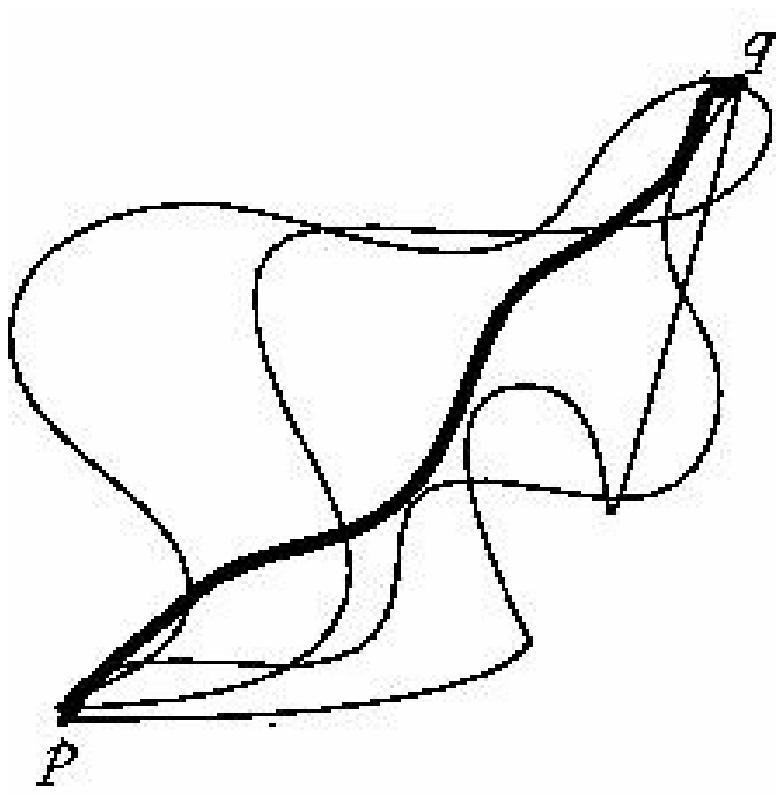}
\end{eqnarray*}
%\begin{figure}
%  \includegraphics[width=2.0in]{FeynmanSum }\\
%  \caption{Path Integral}\label{FeynmanSum}
%\end{figure}
and $$S(\gamma) = \frac{1}{2} \int_0^T |\gamma'(t)|^2 dt.$$
In the picture we stress the classical (Euler-Lagrange) path minimizing $S$.

Moreover, if we think of $|\phi_t\rangle=\phi(q,t)$ as a one parameter family of vectors (\emph{kets}) in $\HH = \Map(M, \complex)$ (usually thought of as a Hilbert space) then we have that the main result of Feynman in this case is that $\phi$ \emph{satisfies the Schr\"odinger equation}.

We can try to extract the formal structure behind formula \ref{Feynman} as follows.

Consider $P_T$ to be a compact 1-dimensional manifold with boundary (namely $P=[0,T]$).  We define the \emph{fields} on a 1-manifold $Y$ to be $$ \FF(Y) =  \Map(Y, M), $$ the \emph{moduli space}\footnote{We will return to the subject of moduli spaces below. They are often orbifolds.} of all maps from $Y$ to $M$. We will divide the boundary of $Y$ into two portions that we will call the \emph{incoming} and \emph{outgoing} boundaries $$\partial Y = \partial_0 Y \coprod \partial_1 Y .$$ As part of the structure we need an \emph{action} map $$S_Y \colon \FF(Y) \to \real$$ which in our case could be given by $$ S(\gamma) = \frac{1}{2} \int_Y |\gamma'|^2. $$

We have the following properties:
\begin{itemize}
\item[i)]  We have restriction maps (forming a \emph{correspondence})
$$\xymatrix{
\ & \FF(Y) \ar[dr]^{\pi_1} \ar[dl]_{\pi_0} & \  \cr
\FF(\partial_0 Y)  & \ & \FF(\partial_1 Y) \cr
}
$$
\item[ii)] Whenever we have $ Y = Y' \cup Y'' $ where $Y' \cap Y'' = \partial_1 Y' = \partial_0 Y''$
$$\xymatrix{
\bullet^0 \ar@{-}[rrr]^{Y'} & & & \bullet^{T_1}  \ar@{-}[rrrrr]^{Y''} & & & & & \bullet^{T_1+T_2} \cr
}$$
then
%$\footnote{We have deliverately depicted the case in which $Y$ is \emph{2-dimensional} rather than 1-dimensional. The whole situation can be generalized, and the 2-dimensional case will be the most important to us in this paper.}
$$ S_Y(\gamma) = S_Y'(\gamma|_{Y'}) + S_Y''(\gamma|_{Y''})$$
\item[iii)]  We have the following pull-back diagram\footnote{The fact that this diagram is cartesian implies that we do have a 1-parameter action on $\HH = \Map (M, \complex) = \Map (\FF(\bullet),\complex)$.}
$$\xymatrix{
\ & \ & \FF(Y) \ar[dr]^{\pi'} \ar[dl]_{\pi''} & \ & \   \cr
\ & \FF(Y') \ar[dl]_{\pi_0} \ar[dr]^{\pi'_1} & \ & \FF(Y'') \ar[dr]^{\pi_1} \ar[dl]_{\pi''_0} & \  \cr
\FF(\partial_0 Y) & \  & \FF(Y' \cap Y'') & \ & \FF(\partial_1 Y'') \cr
}
$$

\item[iv)] The initial ket $|\phi_0\rangle$ evolves along $Y$ according to the formula
\begin{equation}\label{Evolution}
|\phi_T\rangle = (\pi_T)_! (\pi_0^*(|\phi_0\rangle) \cdot e^{-i\hbar S}))
\end{equation}
We will call this \emph{the pull-push evolution formula}. It is the fundamental formula for what follows and requires some clarification.
    \begin{itemize}
    \item $|\phi_0\rangle \in \HH$ can be seen as an element in $\Map(\FF(\partial_0 Y), \complex)$ for $\partial_0 Y =\bullet$ a point and hence $\FF(\partial_0 Y) = \FF(\bullet) =\Map(\bullet,M) \cong M$.
    \item $\pi_0^*(|\phi_0\rangle)$ is an element in $\Map(\FF(Y), \complex)$. In fact when we evaluate at $\gamma \in \FF(Y)$ we get $(\pi_0^*(|\phi_0\rangle))(\gamma) = \phi_0 (\gamma(\partial_0 Y)) = \phi_0(\gamma(0))$.
    \item $(\pi_1)_! \colon \Map(\FF(Y), \complex) \to \Map(\FF(\bullet), \complex)$ is the map that integrates over the fiber of $\pi_1 \colon \FF(Y)\to\FF(\bullet)$ (which in this example is the path space $\PP_q$ and therefore it is given by a path integral). Namely $$ ((\pi_1)_! (\Phi)) (q) = \int_{\PP_q} \Phi(\gamma) \DD\gamma$$
    \item You may want to think of the exponential term as a sort of Chern class for a line bundle over $\FF(Y)$. It causes the integral to become oscillatory, and when $\hbar$ approaches $0$, stationary phase approximation makes the probability that the particle travels the classical (Euler-Lagrange) path approach to $1$. Feynman designed it with this express purpose \cite{FeynmanQED}.
    \item Formula \ref{Evolution} is in fact exactly equivalent to formula \ref{Feynman}.
    \end{itemize}

The algebraic abstract structure that we will extract from this is the following. Define $$\HH_Y \colon = \Map(\FF(Y), \complex)$$ then we have
\begin{itemize}
\item[a)] We will write $\HH$ for $\HH(\bullet)$. To every 0-dimensional manifold we have associated a vector space $\HH$.
\item[b)] To every 1-dimensional manifold (say of length $T$) we have associated a linear operator $$\Psi_T\colon \HH\to\HH$$ $$\Psi_T(\phi_0)=\phi_T.$$
\item[c)] Whenever we glue 1-manifolds, we compose the corresponding linear operators. Namely we have homomorphism from $\real$ to $\GL {}(\HH)$.
\end{itemize}
\end{itemize}

The field theory just described is not topological, for the operators \emph{depend} on the length $T$ of the 1-manifold. In a topological theory the operators are independent on the geometry of the 1-manifold and only depend on their topology (hence we only have two operators, the one associated to \emph{the} interval, and the number associated to the circle).

\end{example}

Here we should mention that in string theory we usually start by assuming that rather than point particles interacting at singular points, we consider extended strings as in the following picture.
\begin{center}
\includegraphics[height=2.5in]{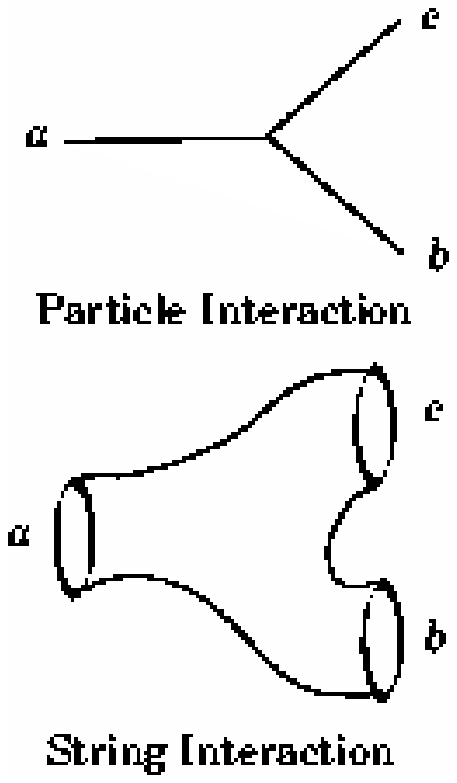}
\end{center}
In the picture we have a particle $a$ scattering in to a pair of particles $b$ and $c$, and the corresponding situation with a string scattering. You should think of this picture as living inside the ambient space time $M$. Notice that the string interaction has \emph{no singularity}.

Traditionally one thinks of $M$ as a smooth manifold, for example in general relativity. Later in this paper we will think instead that the ambient space-time is an orbifold $\Xx$. While a (parameterized) string on a manifold can be modeled by an element of the free loop space $$\gamma \in \LL M = \mathrm{Map} (S^1,M),$$ namely a piecewise smooth map form the circle to $M$, in an orbifold the definition of a loop is more intricate, we will come on this issue later.

Let us remember that a $(n+1)$-dimensional TQFT is a functor $H$ from the category of smooth manifolds and diffeomorphisms to the category of vector spaces that additionally assigns to each cobordism $Y$ between two smooth manifolds $M$ and $N$ (namely $\partial Y = M \amalg -N$), a linear mapping $\Psi_Y \colon H(M) \to H(N)$. Both $H$ and $\Psi$ should be compatible with all the obvious structures including gluing of cobordisms \cite{AtiyahQFT, DijkgraafWitten}. Traditionally whenever $Y$ is boundaryless the map $\Psi_Y\colon \complex \to \complex$ is identified with a number $Z(Y)$. In particular we have that the following properties hold,
\begin{itemize}
    \item $\Psi_\varnothing = \complex$.
    \item $\Psi_{(M \times I)} = \mathbf{1}_{H(M)}$, if $I=[0,1]$.
    \item $H(M \amalg N) \cong H(M) \otimes H(N)$.
    \item $H(-M)\cong H(M)^*$.
    \item $\dim H(M) = Z(M\times S^1)$.
    \item Whenever we have $ Y = Y' \cup Y'' $ where $Y' \cap Y'' = \partial_1 Y' = \partial_0 Y''$ then $$\Psi_Y = \Psi_{Y''} \circ \Psi_{Y'}$$
\end{itemize}
\begin{eqnarray*}
\includegraphics[height=1.3in]{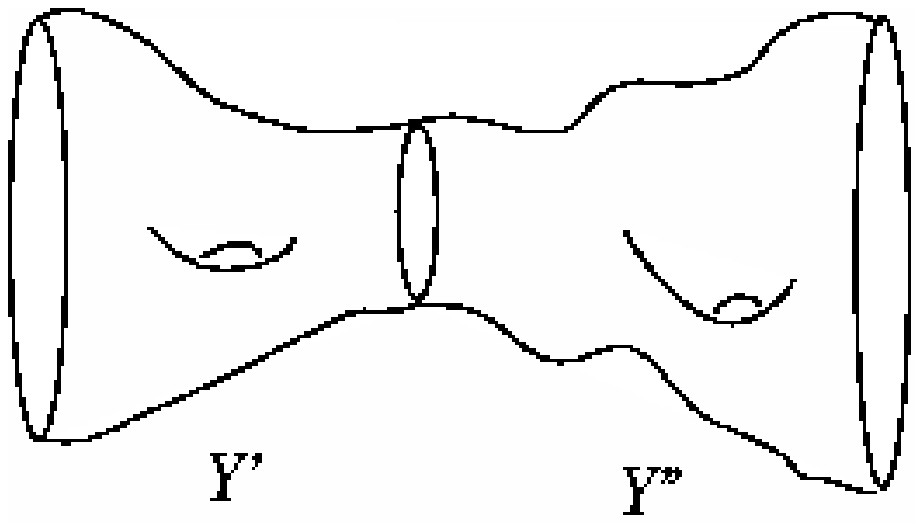}
\end{eqnarray*}

%\begin{figure}
%  \includegraphics[width=2.0in]{gluing }\\
%  \caption{Gluing cobordisms}\label{gluing}
%\end{figure}

    We will also need the definition of a \emph{Frobenius algebra}. A finite-dimensional commutative algebra $A$ over $\complex$ with a unit, together with a linear map $\theta\colon A\to \complex$ such that $(x|y) \mapsto \theta(xy)$ is a nondegenerate bilinear form on $A$ is called a Frobenius algebra. In this case $\theta$ is called a \emph{nonsingular trace} \cite{Abrams, SegalStanford}.

    It is well-known that to have $(1+1)$-dimensional field theory is the same as having a Frobenius algebra. The basic idea is to set $A=H(S^1)$, the product corresponding to the pair of pants $Y_P$, and the trace to a disk $Y_D$ thought of as having one boundary component coming in and none coming out.

\begin{example}
    There is another example afforded to us by Poincar\'{e} duality. This model written $(H^M,\Psi^M,Z^M)_{1+1}\cong(A_M,\theta_M)$ depends only of a fixed oriented compact closed smooth manifold $M$ and lives in dimension $1+1$. Let $\Map^\odot(Y,M)$ be the space of constant maps from $Y$ to $M$.  Clearly  if $Y$ is connected (and non-empty), $\Map^\odot(Y,M)\cong M$ and in fact this last homeomorphism is given by the map $$\ev_y \colon \Map^\odot (Y,M)\to M$$ that evaluates at $y\in Y$. For $Z\subset Y$ we will write $\ev_Z \colon \Map^\odot(Y,M) \to \Map^\odot(Z,M)$ to be the restriction map defined by $\ev_Z(f)=f|_Z$.

In this theory the fields are $$\FF(Y) = \Map^\odot(Y,M),$$ namely the moduli space of constant maps from $Y$ to $M$. We consider $Y$ to be $(1+1)$-dimensional. Notice that $$\Map^\odot(Y,M) = M \times M \times \cdots \times M$$ where the product contains as many copies of $M$ as connected components has $Y$. Consider now the situation in which $Y=P$ a 2-dimensional pair-of-pants (a 2-sphere with three small discs removed) with two incoming boundary components and one outgoing,  and $M$ is an oriented compact closed smooth manifold. Let $a$,$b$ and $c$ be three boundary components $P$ each one diffeomorphic to $S^1$.
$$\xymatrix{
\ & \FF(Y) \ar[dr]^{\pi_1} \ar[dl]_{\pi_0} & \  \cr
\FF(\partial_0 Y)  & \ & \FF(\partial_1 Y) \cr
}
$$
that is to say
\begin{equation}\label{correspondenciaconstante}
\xymatrix{
\ & \Map^\odot(P,M)  \ar[dl]_{\ev_a \times \ev_b} \ar[dr]^{\ev_c} & \  \cr
  \Map^\odot(S^1,M)\times \Map^\odot(S^1,M)  & \ & \Map^\odot(S^1,M)  \cr
}
\end{equation}
which becomes thus
$$\xymatrix{
\ & M \ar[dr]^{=} \ar[dl]_{\triangle} & \  \cr
M\times M  & \ & M \cr
}
$$
and indeed, since that is a smooth correspondence of degree $-d$ we have that
$$\triangle_! = ev_c \circ (\ev_a \times \ev_b)_! \colon H_*(M) \otimes H_*(M) \to H_{*-d}(M)$$
is the induced homomorphism of degree $-d$ in homology. Namely, \emph{the Feynman pull-push evolution for a pair of pants in this field theory is simply the intersection product in homology.}

 We could have used the space $\mathbf{8}$ consisting of the wedge of two copies of $S^1$ instead of $P$ (they are after all homotopy equivalent, we can define $\ev_c$ by choosing a quotient map $c\to\mathbf{8}$ identifying two points of $c$). Notice that by using pairs-of-pants we can recover any compact oriented 2-dimensional cobordism $Y$ which is not boundaryless. In fact by using correspondences we can recover $\Psi^M_Y$ for all $Y$ that has at least one outgoing boundary component. In a sense correspondences encode a big portion of Poincar\'{e} duality this way, the so-called positive boundary sector of the TQFT.

For this model we have,
\begin{itemize}
    \item $A_M= \HH(\bullet) = H_*(M) $ (the homology of $M$ which is graded).
    \item The mapping associated to the pair of pants \begin{equation} \label{product} A_M \otimes A_M \to A_M \end{equation} is the intersection product on the homology of the manifold (and is of degree $-d$).
    \item The trace is defined as $\theta_M\colon A_M = H_*(M) \to  H_*(\bullet) \cong \complex$. The nondegeneracy of the trace is a consequence of Poincar\'{e} duality.
\end{itemize}

It may be instructive to see how the Pontrjagin-Thom construction and the Thom isomorphism can be used to induce the map \ref{product}. That basic idea is to use the \emph{diagonal map} $$\triangle\colon M \to M\times M.$$ $$ m \mapsto (m,m)$$The product on $A_M$ is precisely the Gysin map $\triangle_!$ which can be defined using integration over the fiber, or as follows. It is not hard to verify that the normal bundle $\nu$ of $M= \triangle(M)$ in $M\times M$ is isomorphic to the tangent bundle $TM$ of $M$. Let us write $M_\epsilon$ a small neighborhood of $M$ in $M\times M$, and $M^{TM}$ the Thom space on $TM$. Then we have a natural  map
$$ M\times M \longrightarrow M\times M / (M\times M-M_\epsilon) = M^{TM} $$
which by the use of the Thom isomorphism induces
$$ \triangle_! \colon H_*(M) \otimes H_*(M) \longrightarrow H_{*-d}(M) $$
as desired.
\end{example}

\begin{example}
    This is a famous example due to Chas and Sullivan \cite{ChasSullivan}. Following Cohen and Jones \cite{CohenJones} we do something rather drastic now and let the maps roam free, namely we write the correspondence \ref{correspondenciaconstante} but with the whole mapping spaces rather than just the constant maps.
%\begin{equation}%\label{correspondencianoconstante}
% \LL M\times \LL M = \Map(S^1,M)\times \Map(S^1,M) \stackrel{\ev_a %\times \ev_b}{\longleftarrow} \Map(\mathbf{8},M) %\stackrel{\ev_c}{\rightarrow} \Map(S^1,M) = \LL M,
%\end{equation}
\begin{equation}\label{correspondencianoconstante}
\xymatrix{
\ & \Map(\mathbf{8},M)  \ar[dl]_{\ev_a \times \ev_b} \ar[dr]^{\ev_c} & \  \cr
 (\LL M)^2 = \Map(S^1,M)\times \Map(S^1,M)  & \ & \Map(S^1,M) = \LL M \cr
}
\end{equation}
which \emph{is a degree $-d$ smooth correspondence}. We must replace the pair of pants $P$ for the figure eight space $\mathbf{8}$ in order to ensure that $ \Map(\mathbf{8},M) \to \LL M \times \LL M$ is a \emph{finite} codimension embedding. This in turns implies the existence of the Gysin map $$(\ev_a \times \ev_b)_! \colon H_*(\LL M \times \LL M) \to H_{*-d}( \Map(\mathbf{8},M)).$$

The induced map in homology
$$ \bullet\colon H_*(\LL M) \otimes H_*(\LL M) \to H_{*-d}(\LL M)$$
is called the \emph{Chas-Sullivan} product on the homology of the free loop space of $M$. From the functoriality of correspondences it isn't hard to verify that the product is associative.

Chas and Sullivan proved more, by defining a degree one map $\Delta \colon H_*(\LL M) \to H_{*+1}(\LL M)$ given by $\Delta(\sigma) = \rho_*(\theta \otimes \sigma)$ where $\rho\colon S^1 \times \LL M \to \LL M$ is the evaluation map and $\theta$ is the generator of $H^1(S^1, \integer)$, they proved that $(H_*(M),\bullet, \Delta)$ is a Batalin-Vilkovisky algebra, namely
\begin{itemize}
    \item $(H_{*-d}(M), \bullet)$ is a graded commutative algebra.
    \item $\Delta^2=0$
    \item The bracket $\{\alpha,\beta\} = (-1)^{|\alpha|} \Delta(\alpha \bullet \beta) - (-1)^{|\alpha|} \Delta(\alpha)\bullet \beta - \alpha\bullet\Delta(\beta)$ makes $H_{*-d}(M)$ into a graded Gerstenhaber algebra (namely it is a Lie bracket which is a derivation on each variable).
\end{itemize}
This statement amounts essentially to the construction of $\Psi^{\LL M}_Y$ for all positive boundary genus zero $(1+1)$-dimensional cobordisms $Y$ due to a theorem of Getzler (cf. \cite{Getzler}). The case of positive genus has been studied by Cohen and Godin \cite{CohenGodin}.
\end{example}

\begin{example}
The Gromov-Witten invariants introduced by Ruan in \cite{RuanGW} can be understood in terms of a field theory \cite{Piunikhin}. Now we consider a Riemmann surface $Y=\Sigma_g$ of genus $g$ with $k$ marked points. These marked points will take the place of $\partial_0 Y$ and for simplicity we will not consider outgoing boundary for now.
\begin{eqnarray*}
\includegraphics[height=2.0in]{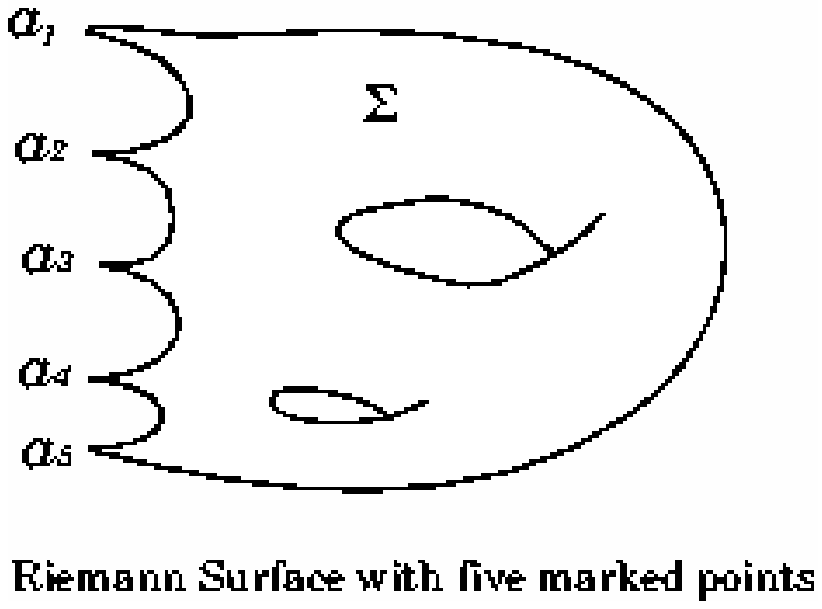}
\end{eqnarray*}
In this (1+1)-dimensional quantum field theory we start by considering a fixed symplectic manifold $(M,\omega)$. The space of fields is given (roughly speaking) by the space of $J$-holomorphic maps on the class $\beta \in H_2(M)$,  $$\FF(Y) = \MM_\Sigma = \Hol_\beta(\Sigma, M)=\{f \in \Hol(\Sigma, M) | f_*[\Sigma] =\beta \},$$
If we denote by $\ev_i\colon \MM_\Sigma \to M$ the evaluation map at $a_i \in \Sigma$, then we have the correspondence diagram
$$\xymatrix{
\ & \MM_\Sigma \ar[dl]^{\times_i \ev_i} \ar[dr] & \  \cr
M^k=\FF(\amalg_i a_i)  & \ & \FF(\varnothing) = \bullet \cr
}
$$
Given $k$ cohomology classes $u_1,\ldots,u_k \in H^*(M)$ we can let them evolve according to Feynman's pull-push formalism to obtain the corresponding \emph{Gromov-Witten invariant} $$ \Phi_{g,\beta,k} (u_1,\ldots,u_k)= \int_{\MM_\Sigma} \ev_1^* u_1 \wedge \ldots \wedge \ev_k^* u_k$$
Here we should mention two important technical points regarding the moduli space $\MM_\Sigma$. Firstly Kontsevich \cite{KontsevichTorus} discovered that the most convenient space for defining this field theory is the moduli space of stable maps (where at most ordinary double points are allowed, and with finite automorphism groups). The moduli space turns out to be an orbifold, not a manifold. We will return to the definition of an orbifold later.

Secondly, the moduli space does not quite have a fundamental class (that we require to do the integration). The problem is that roughly speaking $\MM$ is given as the intersection of two submanifolds (equations) $N_1$ and $N_2$ of a larger manifold $V$ (taking only two is possible by using the diagonal map trick, namely $N_1 \cap \ldots \cap N_r = (N_1 \times \cdots \times N_r) \cap \bigtriangleup(V^r)$). Often this intersection is not transversal. Therefore rather than a tangent we have a \emph{virtual} tangent bundle (in $K$-theory) $$ [T\MM]^\virt = [TN_1]|_\MM + [TN_2]|_\MM
 - [TV]|_\MM$$ whose orientation (in cohomology, K-theory, complex cobordism) is called the \emph{virtual fundamental class} $[\MM]^\virt$. The corrected formula for the Gromov-Witten invariants is then $$ \Phi_{g,\beta,k} (u_1,\ldots,u_k)= \int_{[\MM_\Sigma]^\virt} \ev_1^* u_1 \wedge \ldots \wedge \ev_k^* u_k.$$
\end{example}

\begin{example} Floer theory is also a quantum field theory. Now we consider $Y=\Sigma_{g,k}$ to be a genus $g$ Riemann surface with $k$ small discs removed.
\begin{eqnarray*}
\includegraphics[height=2.0in]{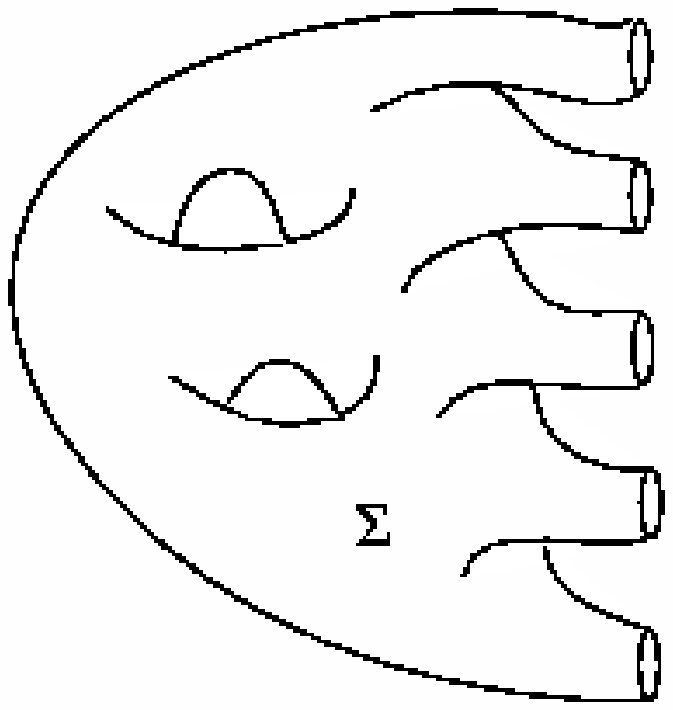}
\end{eqnarray*}
The fields are again holomorphic mappings $\FF(Y) = \MM_\Sigma$.
$$\xymatrix{
\ & \MM_\Sigma \ar[dl]^{\times_i \ev_i} \ar[dr] & \  \cr
\FF(\varnothing) = \bullet  & \ & (\LL M)^k = \FF(\partial_1 \Sigma) \cr
}
$$

In this case rather than simply considering the homology of $\LL M$ we consider its semi-infinite (co)homology. This means that we consider the homology of cycles that are half-dimensional in $\LL M$. The semi-infinite (co)homology $H^{\mathrm{si}}_*(\LL M)$ is also known as the \emph{Floer} (co)homology $HF_*(M)$.

\emph{Cohen's conjecture} states that the quantum field theory of Chas-Sullivan on a manifold $M$ is isomorphic to the Floer quantum field theory of $T^* M$ (which is always symplectic).

\end{example}

\section{Group Actions}

Given a space $M$ we often want to study all its self-transformations that preserve some of its properties. Often such transformations are called \emph{symmetries} and often they are also called \emph{automorphisms}.

\begin{example} Consider a triangle $T$ as a subset of $\real^2$. We may ask how many mappings $g \colon T \to T$ there are with the property $$ d(x,y) = d(g(x),g(y))$$ for every pair of points in the triangle, where $d$ denotes the usual distance. Such a map is called an \emph{isometry} of the triangle.

The answer of course depends very much on the triangle.

      \begin{itemize}

          \item If the triangle is scalene only the identity is an isometry of $T$.

          \item If the triangle is isosceles then there are two such isometries.

          \item If the triangle is equilateral there are six isometries of $T$.

      \end{itemize}

      This can be verified by noticing that an isometry is completely determined by its restriction to the vertices.

\end{example}

       Here, as we all know, we can take a remarkable conceptual leap: \emph{we decide to remember how the different symmetries interact} rather than the symmetries themselves. For this we observe that

   \begin{itemize}

      \item If $g$ and $h$ are symmetries of $T$ so is $g\circ h= gh$.

      \item $(gh)k=g(hk)$

      \item There is always the identity symmetry $1_T$.

      \item Given a symmetry $g$ there is another symmetry $k$ such that $gk=kg=1_T$.

   \end{itemize}

   This motivates the definition of (abstract) \textbf{group} \cite{AlgebraLang}. A group is a set of things, together with a composition law that satisfies all the previous axioms. We say for example that the isometries of $T$ form a group.

    Once we have this definition we end up with groups that are (at first) not naturally the symmetries of anything. For example, the fundamental group of a space $X$ is at first an abstract group formed with homotopy classes of paths. In this case it may come as a surprise to learn that $\pi_1(X)$ in fact acts as some sort of symmetry, namely as \emph{deck} transformations of the universal cover $M=\widetilde{X}$.  It is often important to realize that an (abstract) group is indeed a group of transformations of some space $M$.

 \begin{definition}  We say that the group $G$ \emph{acts on} the object $M$ if we are given a homomorphism $$ \psi \colon G \to \Aut(M),$$  Namely, for every $g\in G$ and every $m \in M$ we have

   \begin{itemize}
      \item $ mg = \psi(g)(m) \in M$ such that
      \item $m1_M = m$
      \item $(mg)h = m(gh)$
   \end{itemize}

\end{definition}

\begin{definition} We say that the group $G$ acts \emph{effectively} on the object $M$ if $ \psi \colon G \to \Aut(M)$ is injective, namely for all $g \in G, g \neq 1$ there is an $m\in M$ so that $mg \neq m$.
\end{definition}

\begin{definition}  The equivalence relation \emph{induced} by the action of $G$ on $M$ is the relation generated by $$x \sim xg.$$ The quotient $M/\sim$ is also written $$M/G.$$ The equivalence classes of this relation are called the \emph{orbits} of the action. They are written $$ [ m ] =m\cdot G = \{ mg | g\in G\}.$$ If there is only one equivalence class (orbit) for the action we say that $G$ acts \emph{transitively} on $M$.
\end{definition}

\begin{definition} The stabilizer subgroup of $m \in M$ is $$G_m = \{ g\in G | mg = m\}.$$ Notice that even effective actions often have nontrivial stabilizers.
\end{definition}

\begin{proposition} If $G$ acts on $M$ then $$ G / G_m \simeq m\cdot G$$ as \emph{sets}.
\end{proposition}

\begin{example} Let $M=P$ be the set of all lines in $\real^3$ containing the origin. Then the group of all linear automorphisms of $\real^3$, $G=\GL{3}(\real)$ acts on $M$. Let $m\in M$ be the $x$ axis. Then it is not hard to see that $G_m = \GL{2}(\real)$ and therefore $$ P = \GL{3}(\real) / \GL{2}(\real)$$
\end{example}

We write $$p\colon M \to M/G$$ for the mapping $$ m \mapsto [m]$$

If $M$ is a topological space and $G$ acts on $M$ then we can put a natural topology on $M/G$, namely a subset $U$ of $M/G$ is declared to be open if and only if $p^{-1}(U)$ is open in $M$.

\begin{example} $\widetilde{X} / \pi_1(X) \simeq X$.
\end{example}

There are quotients in the category of sets, and also in the category of topological spaces.

But the category of smooth manifolds is quite unlike the category of sets or of topological spaces (for manifolds have structure sheafs).

\section{Orbifolds.}

Let $M=T^2 = S^1 \times S^1$ be a two-dimensional torus, and let $G= \integer_2$ be the \emph{finite} subgroup of diffeomorphisms of $M$ given by the action
$$ (z,w) \mapsto (\bar{z}, \bar{w})$$

\begin{eqnarray*}
\includegraphics[height=2.0in]{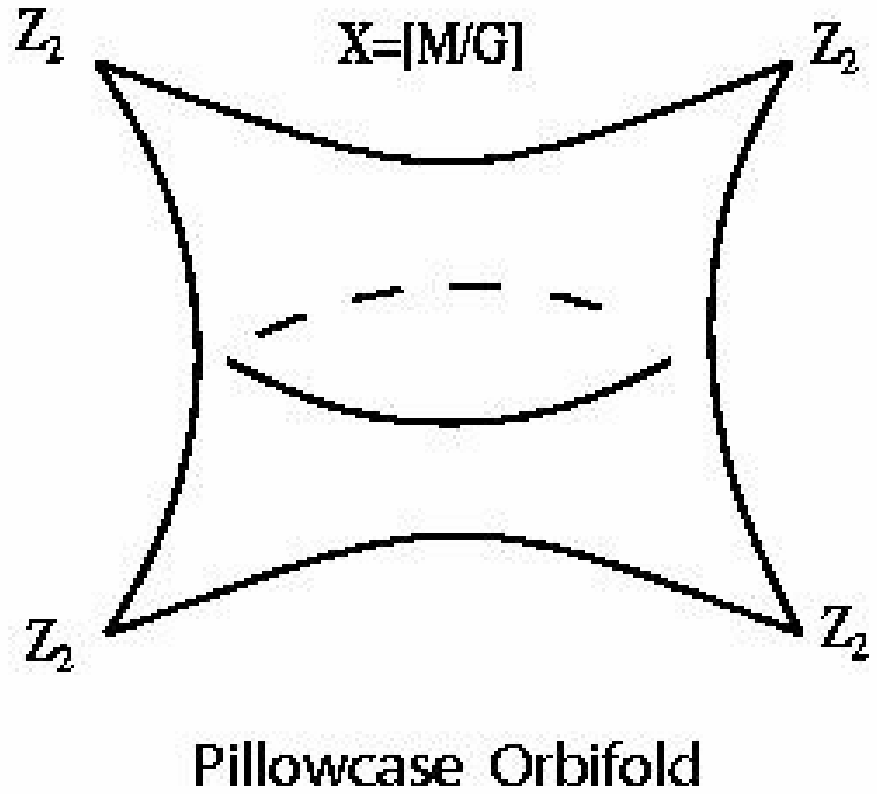}
\end{eqnarray*}

\begin{example} \label{ex. pillowcase} Show that while the quotient space $X=M/G$ is topologically a sphere it is impossible to put a smooth structure on $X$ so that the quotient map $M \To X$ will become smooth. It is in this sense that we say that $X$ is not a smooth manifold.
\end{example}

What will enlarge the category of smooth manifolds to a bigger category is called the category of \emph{orbifolds}.  Once we do this, when we consider the orbifolds $\Mm$ and $\Xx$ then the natural orbifold morphism $\Mm \To \Xx$ becomes smooth.

While the orbifold $\Mm$ contains exactly the same amount of information as $M$ the orbifold $\Xx=[M/G]$ (known as a pillowcase) contains more information that the quotient space $X=M/G$.  For instance $\Xx$ remembers that the action had 4 fixed points each with stabilizer $G$. It remembers in fact the stabilizer of every point, and how these stabilizers fit together.
On the other hand $\Xx$ does not remember neither the manifold $M$ nor the group $G$.  In fact if we define $N$ to be two disjoint copies of $M$ and $H = G \times G$ to act on $M$ by letting $G\times1$ act by complex conjugation on both copies as before, and $1\times G$ act by swapping the copies then $$\Xx = [M/G] \cong  [N/H].$$

 Not every orbifold can be obtained from a finite group acting on a manifold. An orbifold is always \emph{locally} the quotient of a manifold by a finite group but this may fail globally.

For example consider the teardrop:

\begin{eqnarray*}
\includegraphics[height=2.0in]{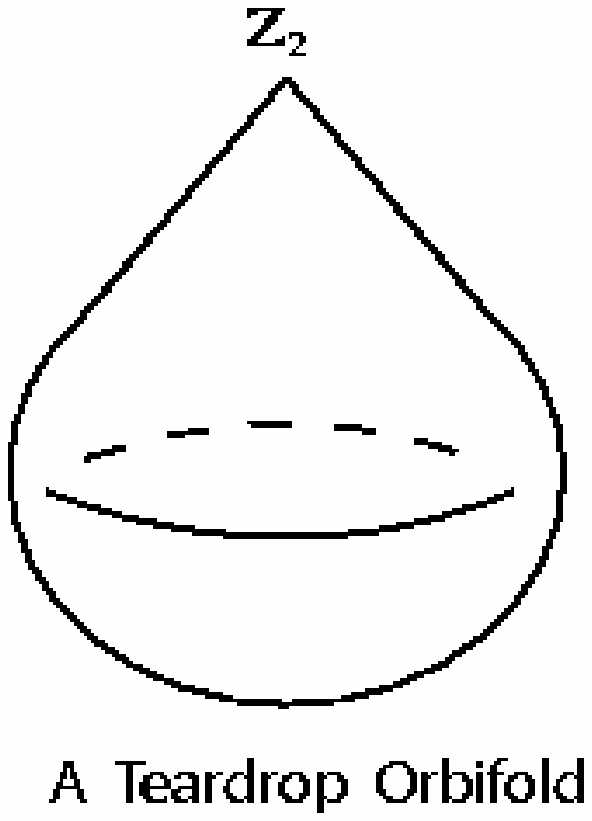}
\end{eqnarray*}

This orbifold may be obtained by gluing two global quotients. Consider the orbifold $\Xx_1 = [\complex / \integer_2]$ where $\integer_2$ acts by the holomorphic automorphism $z \mapsto -z$. Let $\Xx_2 = \complex$ simply be the complex line. Then we have in the category of orbifolds a diagram of inclusions
$$ \Xx_1 \longleftarrow \complex^* \longrightarrow \Xx_2 $$
and therefore we can glue $\Xx_1$ and $\Xx_2$ along $\complex^*$ to obtaining the teardrop $\Xx$.

\begin{example} Prove that it is not possible to obtain the teardrop as a global quotient by a finite group.
\end{example}

There are several
definitions of the concept of an orbifold.  The first one due to
Satake \cite{Satake} was written using the so-called orbifold
atlases, unfortunately quite a few concepts are a bit cumbersome using
this definition.  We opt to think of an orbifold as a certain kind
of category following Grothendieck, Haefliger
and Moerdijk \cite{Moerdijk}.

For this reason we will start reviewing the basic facts about category
theory. Category theory was discover by Eilenberg and
MacLane in the 50's  \cite{MacLane} and
ever since has pervaded all fields of mathematics.

You may want to think of the category of sets as you read the following definition. The objects of the category of sets are all sets and the arrows are all mappings between them. You may also want to think of an object as a sort of dot and an arrow as something with a direction joining the dots.

\begin{definition} A \textbf{category} consists of:
\begin{itemize}
\item A class $\obj(\CC)$, \index{objects, class of}that we will
denote by $\CC_0$, of objects of $\CC$.

\item A class $\arr(\CC)$, that we will denote by $\CC_1$, of
arrows\index{arrows, class of} of $\CC$. For each pair of objects
$a$ and $b$ the class of all arrows from $a$ to $b$ is denoted by
$\CC(a,b)$.

\item Two assignments $s_{\CC},t_{\CC}:\arr(\CC)\rightarrow \obj(\CC)$
called source and target respectively.

\item \emph{Unit}. An assignment $u_{\CC}:\obj(\CC)\rightarrow
\arr(\CC)$ such that:

\[
s_{\CC}( u_{\CC}(a))=t_{\CC}( u_{\CC}(a))=a,
\]
for every $a\in \obj(\CC)$.

 \item \emph{Composition Law}. For each triple $a$, $b$  and
$c$ of objects of $\CC$ an assignment
$m_{(a,b,c)}:\CC(a,b)\times\CC(b,c)\rightarrow\CC(a,c) $, where
its image on $(\ag,\bg)\in \CC(a,b)\times \CC(b,c)$ well be
denoted by $\bg\circ \ag$, satisfying the following properties:
\begin{enumerate}

\item For every $a\in \obj(\CC)$
\[
s_{\CC}( u_{\CC}(a))=t_{\CC}( u_{\CC}(a))=a,
\]
\[
\xymatrix{\obj(\CC) \ar[d]_{u}\ar[r]^{u}\ar[dr]^{Id}    & \arr(\CC) \ar[d]^{s} \\
          \arr(\CC)\ar[r]^{t}                           & \obj(\CC)}
\]
in other words the source and target of $u_{\CC}(a)=a$ for every
$a$.

\item \emph{Associativity}. For all $\ag,\bg,\gamma\in\arr(\CC)$
it holds that $\ag\circ(\bg\circ\gamma)=(\ag\circ\bg)\circ\gamma$,
formally for every elements $a,b,c,d$ fixed in $\obj(\CC)$ we have
\[
m_{(a,c,d)}\circ(m_{(a,b,c)}\times
Id_{\CC(c,d)})=m_{(a,b,d)}(Id_{\CC(a,b)}\times m_{(b,c,d)}),
\]

\[
\xymatrix{\CC(a,b)\times\CC(b,c)\times\CC(c,d)\ar[d]_{Id_{\CC(a,b)} \times m_{b,c,d}}\ar[rr]^{m_{a,b,c}\times
Id_{\CC(c,d)}}       &    &\CC(a,c)\times\CC(c,d)\ar[d]^{m_{(a,c,d)}}\\
\CC(a,b)\times\CC(b,d)\ar[rr]^{m_{(a,b,d)}}
      &  &\CC(a,d)}
\]

\item \emph{Unity}. For every $a,b \in \obj(\CC)$ and  $\ag\in
\CC(a,b) $ $\ag=u_{\CC}(b)\circ\ag=\ag\circ u_{\CC}(a)$ holds,
formally
\[
m_{(a,b,b)}(\ag,u_{\CC}( b))=m_{(a,a,b)}(u_{\CC}(a),\ag)=\ag.
\]
\end{enumerate}

\end{itemize}

\end{definition}

\begin{example}
Let us define \Sets \ the category with objects the class of all
spaces (proper class) and arrows the class of function of sets. The
unity of this category assigns to each set $X$ the usual identity
function of sets over $X$ and the function $m_{\CC}$ the
composition of functions, when it is defined.
\end{example}

\begin{example} The category \Abel \ the subclass of \Sets \ whose objects are all
abelian groups and arrows the class of morphism of groups with the
same unity and rule of composition as \Sets. In he same manner are
defined the categories \Modules,  \Rings, \Groups, etc.
\end{example}

\begin{example} The category \Top\ of topological spaces and continuous
functions.
\end{example}

\begin{example}
Let us consider the category $\Corr$ of correspondences \cite{CohenVoronov} whose objects are topological spaces and whose arrows (from $X$ to $Y$) are diagrams of continuous mappings of the form
$$\xymatrix{
\ & Z  \ar[dl]_{\alpha} \ar[dr]^{\beta} & \  \cr
 X  & \ & Y \cr
}$$
for $Z$ some topological space. We define the composition of arrows by $$(X \stackrel{\alpha}\leftarrow V \stackrel{\beta}{\rightarrow} Y)\circ(Y \stackrel{\gamma}{\leftarrow} W \stackrel{\delta}{\rightarrow} Z) = X \stackrel{\alpha}{\leftarrow} U \stackrel{\delta}{\rightarrow} Z$$ where $U$ is defined as the fiber product $$U = V \times_Y W =\{ (v,w) | \beta(v)=\gamma(w)\}.$$

Observe that the ordinary category of topological spaces can be embedded as a subcategory of $\Corr$ since a continuous map $f\colon X\to Y$ can be interpreted as the correspondence $$ X \stackrel{\pi_X}{\leftarrow} \GG_f \stackrel{\pi_Y}{\rightarrow} Y,$$ where $\GG_f=\{(x,y)|y=f(x)\}$ is the graph of $f$. This is functorial for we have $$\GG_f \times_Y \GG_h = \GG_{h \circ f}.$$

Unfortunately homology is not a functor from $\Corr$ to graded abelian groups. Nevertheless suppose that we have a correspondence $X \stackrel{\alpha}{\leftarrow} Z \stackrel{\beta}{\rightarrow} Y$ where \begin{itemize}
    \item $X$, $Y$ and $Z$ are manifolds (possibly infinite dimensional).
    \item $\alpha$ is a regular embedding of finite codimension $d$.
\end{itemize}
In this case we say that $X \stackrel{\alpha}{\leftarrow} Z \stackrel{\beta}{\rightarrow} Y$ is a smooth correspondence of degree $-d$. In any case using the Gysin map we can produce the composition $$H_*(X) \stackrel{\alpha_!}{\rightarrow} H_{*-d}(Z) \stackrel{\beta_*}{\rightarrow} H_{*-d}(Y)$$ which is the induced homomorphism of degree $-d$ in homology.
\end{example}

\begin{definition} A \emph{Groupoid} is a category in which each arrow has an
inverse, namely for each pair $ a,b\in\obj\CC$ and each
$\ag\in\CC(a,b)$ there exist an arrow $\ag^{-1}\in\CC(b,a)$ in
such a way that $\ag^{-1}\circ\ag=u(a)$ y $\ag\circ\ag^{-1}=u(b)$.
In this case we will denote by $i:\CC(a,b)\rightarrow \CC(b,a)$
the map that assigns to each arrows its inverse.
\end{definition}

\begin{example} Let $G$ be a group acting on a set $M$. Let $G\ltimes M$ be the
groupoid whose objects are the set $\M$, and arrows $g:x\rightarrow y$
 such that $y=gx$, this set can be seen as the set
$G\times M$. Here the composition is defined of natural manner
$gg':x\rightarrow z$ where $g':x\rightarrow y$ and $g:y\rightarrow
z$. For each object $x$ the unit map associates the unit $e$ of
$G$. The structure maps are defined in the obvious way as $s:G\times
M\rightarrow M$ the projection and $t:G\times M\rightarrow M$ the
action.
\end{example}

\begin{example} Smooth manifolds provide a natural source of groupoids.
Let $M$ be a smooth manifold. It is well known that a smooth manifold is a pair $(M,\UU)$ of a (Hausdorff, paracompact) topological  space $M$ together with an atlas $\UU=(U_i)_{i\in I}$, and is only by abuse of notation that we speak of a manifold $M$. In fact a smooth manifold is actually an equivalence class of a pair $ [M,\UU]$ where we say that $(M,\UU_1) \sim (M,\UU_2)$ if and only if there is a common refinement $(M,\UU_3)$ of the atlas. We can say this in a slightly different way that will be easier to generalize to the case of orbifolds. To have a pair $(M,\UU)$ is the same thing as to have a small topological category $\Mm_\UU$ defined as follows.
\begin{itemize}
   \item Objects: Pairs $(m,i)$ so that $m \in U_i$. We endow the space of objects with the topology $$ \coprod_i U_i.$$
   \item Arrows: Triples $(m,i,j)$ so that $m \in U_i \cap U_j =U_{ij}$. An arrow acts according to the following diagram. $$(x,i)\toparrow{(x,i,j)}(x,j).$$
   \item The composition of arrows is given by $$(x,i,j)\circ(x,j,k)=(x,i,k)$$ The topology of the space of arrows in this case is $$ \coprod_{(i,j)} U_{ij}.$$
  \end{itemize}

The category $\Mm$ is actually a {\textbf{groupoid}},  in fact $$(x,i,j) \circ (x,j,i) =(x,i,i) = Id_{(x,i)}.$$ We will therefore define a manifold to be the equivalence class of the groupoid $\Mm_\UU$ by an equivalence relation called Morita equivalence (that will amount exactly to the equivalence of atlases in this case).

\end{example}

\subsection{Homotopies.}

 The category $\Top$ of topological spaces with continuous mappings has an interesting additional structure. Homotopies of smooth mappings. This endows $\Top(X,Y)$ with the structure of a category. We will call a category with this additional structure a \emph{bicategory}.

The category $\Cat$ of all categories is also a bicategory. Let us define the homotopies between functors.
Let $F$ and $D$ functors from $\CC$ to $\BB$, a \emph{homotopy} of functors is a functor $H \colon \CC \times \II \to \BB$ where $\II$ is a category with two objects and one arrow going between them, and the restrictions of $H$ to the two copies of $\CC$ above, coincide with $F$ and $D$ respectively. The reader can verify that to have a homotopy between functors is the same as having a natural transformation.

\begin{definition} A natural transformation of functors is a map $\Phi: \CC_0\rightarrow \BB_1$ in such a
way that

\begin{itemize}
\item For every $a\in\CC_0,\Phi(a)\in \BB(F(a),D(a))$, and

\item For each $\ag\in\CC(a,b)$
\[
\Phi(b)\circ F(\ag)=D(\ag)\circ\Phi(a)
\]
\[
\xymatrix{F(a)\ar[r]^{F(\ag)}\ar[d]_{\Phi(a)}
&F(b)\ar[d]^{\Phi(b)}\\
          D(a)\ar[r]^{D(\ag)}  &D(b) }
\]
\end{itemize}

\end{definition}

\section{Groupoids}

\begin{definition}
A Lie groupoid $\Gg$  is a category in which every
morphism is invertible such that $\Gg_0$ and $\Gg_1$, the sets of
objects and morphism respectively, are \emph{smooth manifolds}.  We will denote the structure maps by:
      $$\xymatrix{
        \Gg_1 \timests \Gg_1 \ar[r]^{m} & \Gg_1 \ar[r]^i &
        \Gg_1 \ar@<.5ex>[r]^s \ar@<-.5ex>[r]_t & \Gg_0 \ar[r]^e & \Gg_1
      }$$
where $s$ and $t$ are the source and the target maps, $m$ is the composition (we can compose two arrows whenever the
target of the first equals the source of the second), $i$ gives us
the inverse arrow, and $e$ assigns the identity arrow to every
object. We will assume that all the structure maps are smooth maps.  We also require the
maps $s$ and $t$ to be submersions, so that $\Gg_1 \timests
\Gg_1 $ is also a manifold.
\end{definition}

\begin{definition} The \emph{stabilizer} $\Gg_x$ of a groupoid $\Gg$ on $x \in \Gg_0$ is the set of arrows whose source and target are both $x$. Notice that $\Gg_x$ is a group \end{definition}

\begin{definition} A topological (Lie) groupoid is called {\it \'{e}tale} if the
source and target maps $s$ and $t$ are local homeomorphisms (local
diffeomorphisms).

For an \'{e}tale groupoid we will mean a
topological \'{e}tale groupoid.

We will always denote groupoids
by letters of the type $\Gg,\Hh,\Ss$.

We will also assume that the  anchor map $(s,t):
\Gg_1 \to \Gg_0\times \Gg_0$ is proper, groupoids with this
property are called {\it proper groupoids}.  This will force all stabilizers to be finite.

\end{definition}

\begin{definition} A morphism of groupoids $\Psi: \Hh \to \Gg$ is a pair of maps
$\Psi_i: \Hh_i \to \Gg_i$ $i=0,1$ such that they commute with the
structure maps. The maps $\Psi_i$ will be required to be smooth.

The morphism $\Psi$ is called {\it Morita} if the following
square is a cartesian square .
\begin{eqnarray}
 \xymatrix{
         \Hh_1 \ar[r]^{\Psi_1} \ar[d]_{(s,t)} & \Gg_1 \ar[d]^{(s,t)} \\
         \Hh_0 \times \Hh_0 \ar[r]^{\Psi_0 \times \Psi_0} & \Gg_0 \times \Gg_0
         } \label{Moritasquare}
\end{eqnarray}
and if  $s \circ \pi_2:\Hh_0
\: {}_{\Psi_0} \! \times_{t} \Gg_1 \to \Gg_0$ is an open
surjection.

Two groupoids $\Gg$ and $\Hh$ are Morita equivalent if
there exist another groupoid $\Kk$ with Morita morphisms $\Gg
\stackrel{\simeq}{\leftarrow} \Kk \stackrel{\simeq}{\to} \Hh$.
\end{definition}

 A theorem of Moerdijk \cite{Moerdijk} states that the category of orbifolds is equivalent to a quotient category of the category of proper \'{e}tale groupoids after formally inverting the Morita morphisms.

Whenever we write orbifold, we will choose a proper
\'{e}tale smooth groupoid representing it (up to Morita
equivalence.)

\begin{example}
   Consider again the pillowcase (as in example \ref{ex. pillowcase}). Define the following groupoids.

   \begin{itemize}
      \item The groupoid $\Gg$ whose space of objects are elements $m \in M$ with the topology of $M$, and whose space of arrows is the set of pairs $(m,g)$ with the topology of $M\times G$. We have the diagrams $$m\toparrow{(m,g)} mg$$ and the composition law $$(m,g) \circ (mg, h) = (m,gh).$$

      \item Similarly we define the groupoid $\Hh$ using the action of $H$ in $N$ with objects $n\in N$ and arrows $(n,h)\in N\times H$.
   \end{itemize}

The orbifold $\Xx$ is the equivalence class of the groupoid $\Gg$. Since there is a Morita morphism $\Hh \to \Gg$, we can say also that $\Xx$ is the equivalence class of $\Hh$. By abuse of notation we will often say that $\Gg$ is an orbifold when we really mean that its equivalence class is the orbifold.

\end{example}

\begin{example}
   More generally, let $M$ be a smooth manifold and $G \subset {\mathrm{Diff}}(M)$ be a finite group acting on it.

\begin{itemize}

\item  We say that the orbifold $[M/G]$ is the equivalence class of the groupoid $\Xx$ with objects $m \in M$ and arrows $(m,g) \in M\times G$.

\item  We can define another groupoid representing the same orbifold as follows. Take a contractible open cover $\UU=\{U_i\}_{i \in I}$
of $M$ such that all the finite intersections of the cover are either contractible or empty, and with
the property that for any $g \in G$ and any $i \in I$ there exists $j \in I$ so that $U_ig=U_j$. Define
$\Gg_0$ as the disjoint union of the $U_i$'s with $\Gg_0 \stackrel{\rho}{\to} M=\Xx_0$ the natural map.
Take $\Gg_1$ as the pullback square
  $$\xymatrix{ \Gg_1 \ar[r] \ar[d] & M \times G \ar[d]^{\source \times \target} \\
             \Gg_0 \times \Gg_0 \ar[r]^{\rho \times \rho}& M \times M}$$
where $\source(m,g) =m$ and $\target(m,g) = mg$.
From the construction of $\Gg$ we see that we can think of $\Gg_1$ as the disjoint union of all the intersections
of two sets on the base times the group $G$, i.e.
$$\Gg_1 = \left( \bigsqcup_{(i,j) \in I\times I} U_i \cap U_j \right) \times G$$
where the arrows in $U_i \cap U_j \times\{g\}$ start in $U_i|_{U_j}$ and end in $(U_j|_{U_i})g$.  This defines a proper \'{e}tale Leray groupoid $\Gg$ and
by definition it is Morita equivalent to $\Xx$.
\end{itemize}
\end{example}

\section{Moduli Spaces}

Moduli spaces are often given by orbifolds. Moduli spaces are ``spaces'' that contain the universal family of objects of certain kind. If $\Xx$ is the moduli space of objects of certain kind we want $$\mathrm{Maps}(S,\Xx)$$ to classify families of objects of this kind over $S$. This is akin to the situation in topology in which we represent, for example $n$-dimensional vector bundles over $M$ \emph{up to isomorphism} by \emph{homotopy classes} of maps to a certain universal space $B\U{n}$. Remember that $B\U{n} = \Gr{n}{\complex^\infty}$. Moduli spaces are often not spaces at all but rather \emph{orbifolds}.

\begin{example} Let us consider the \textbf{moduli space of triangles} $\TT$.  We identify an Euclidean triangle $T$ with a triple $$T=(a,b,c)$$ satisfying the triangle inequalities $$ a+b > c,$$ $$b+c > a,$$ and $$c+a>b.$$ The \emph{set} $M$ of all such $T$ is diffeormorphic to $$M \approx\Delta \times \real^+.$$ It is a positive cone over an equilateral triangle (of triangles of fixed perimeter $a+b+c$) that we denote by $\Delta$.

\begin{eqnarray*}
\includegraphics[height=3.5in]{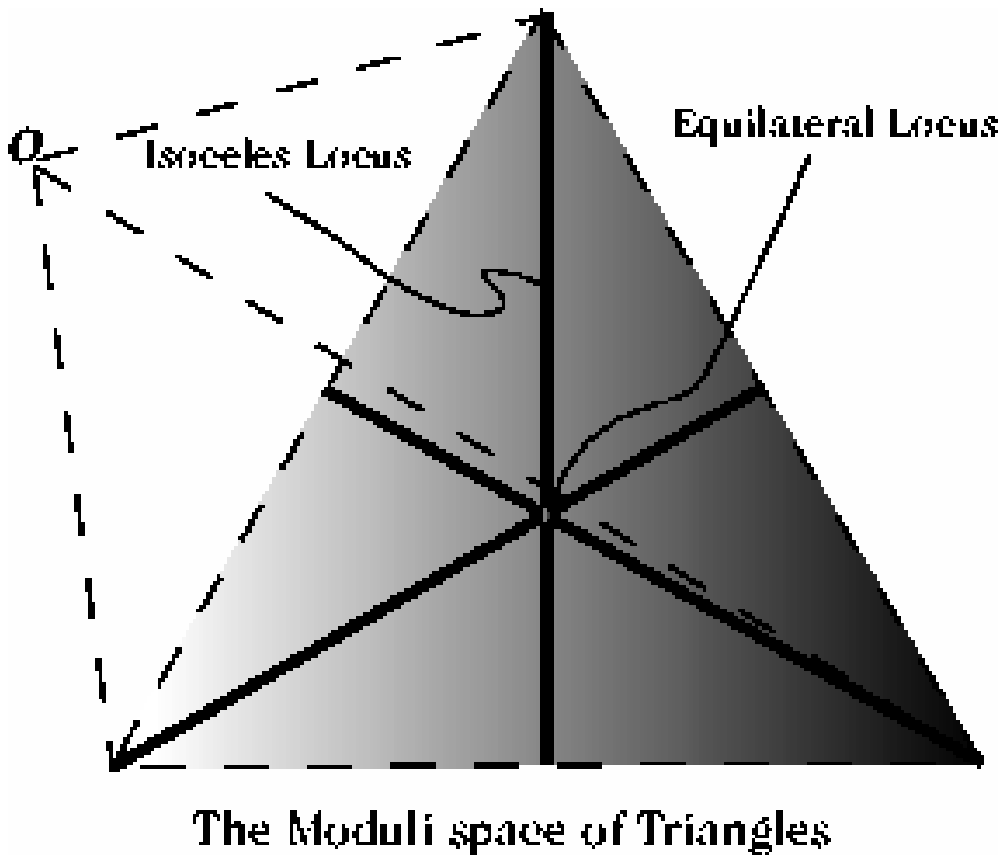}
\end{eqnarray*}

 The is a natural action of $\frak{S}_3$ on $M$ by multiplication of the corresponding permutation matrix. The moduli orbifold of triangles is $$\TT = [ M / \frak{S}_3 ]$$ Now the class of \emph{smooth} families of triangles over the circle $S=S^1$ is now endowed naturally with the structure of an orbifold: $$\TT^S = [ P / \frak{S}_3 ]$$ where $P$ is the family of paths $I=[0,1] \to M$ so that $$\gamma(1) = \gamma(0) \cdot g$$ for some $ g \in  \frak{S}_3 $. This is what we have called the \emph{loop orbifold} \cite{LupercioUribeLoopGroupoid, OrbifoldStringTopology}. We will come back to this later.

\end{example}

\section{Almost Free Lie Group Actions.}
 We will suppose now that $K$ is a  Lie group. Let $M$ be a smooth manifold in which $K$ is acting. We say that $M$ is a $K$-manifold.

 A map $\phi\colon M \to N$ between $K$-manifolds is said to be \emph{equivariant} if $$\phi(xg) =\phi(x)g.$$ We say that a vector bundle $E\longrightarrow M$ is a $K$-vector bundle if $K$ acts linearly on the fibers and the projection map is equivariant.

 Stabilizers $K_m$ of Lie group actions are closed subgroups and hence Lie groups. Stabilizers of points in the same orbit are conjugate to each other: $$K_{mg} = g^{-1} K_m g$$ The conjugacy class of a subgroup $H$ will be written $(H)$. Hence $(K_m)$ only depends on the orbit of $m$ and not on $m$. Given $m\in M$ he map $$f_m \colon K/K_m \longrightarrow M$$ given by $$f_m(\overline{g}) = mg,$$ is an injective immersion. It does not follow that  $m\cdot K \subseteq M$ is a submanifold. Just think of a torus with an irrational flow. Nonetheless, of course, if $K$ is \emph{compact} then $m\cdot K \subseteq M$ is always a submanifold. If $K$ is compact $M/K$ is Hausdorff and $p \colon M \to M/K$ is proper and closed. So, from now on we shall suppose that $K$ is compact. Fix $m\in M$ and let $$V_m = T_m M / T_m (m K).$$ Notice that for $g\in K_m$ we have $$d_m g \colon T_m M \longrightarrow T_{mg} M = T_m M$$ Therefore $$K_m \longrightarrow \Aut(V_m).$$ Also $K_m$ acts freely on $K \times V_m$, by $h(g,v)=(gh^{-1},hv)$. This defines a vector bundle  $K \times_{K_m} V_m\longrightarrow K/K_m$.

\begin{theorem}[The Slice Theorem (Koszul 1953) \cite{Koszul}]  There exists an equivariant diffeomorphism from an equivariant open neighborhood of the zero section of $K \times_{K_m} V_m\longrightarrow K/K_m$ to an open neighborhood of $mK \subseteq M$, sending the zero section to $mK$ by $f_m$.
\end{theorem}

The union of all the orbits of a given type is a submanifold of $M$. If $M$ is compact there are only finitely many orbit types.

From now on we will suppose that all $K_m$ are finite, and that $M/K$ is connected. Then there exists a finite group $G$ so that the set of points in $M$ with stabilizers conjugate to $G$ (denoted by  $M_{(G)}$) is open and dense in $M$. (Prove it by induction over the dimension of the manifold $M$, and consider the sphere bundle of the neighborhoods of the theorem.)

 If $K$ is a compact Lie group acting on $M$,
and each stabilizer $K_x$ is finite, then $K\ltimes M$ is an
orbifold groupoid. Observe that the slice theorem for compact
group actions gives for each point $x$ a "slice" $V_x\subseteq M$
for which the action defines a diffeomorphism $K\times
_{K_x}V_x\hookrightarrow M$ onto a saturated open neighborhood
$U_x$ of $x$. Then $K_x\ltimes V_x$ is an \'etale groupoid which is
Morita equivalent to $K\ltimes U_x$. Patching these \'etale
groupoids together for sufficiently many slices $V_x$ yields an
\'etale groupoid Morita equivalent to $K\ltimes M$ \cite{AdemRuan}.

\begin{definition} A orbivector bundle over $\Xx$ is a pair $(E,\theta)$ where $E$ is an ordinary vector bundle over $\Xx_0$ and $\theta$ is an isomorphism $s^*E \cong t^*E$ (here we are choosing a representative of the Morita class) \end{definition}

\begin{example}This recovers the usual definition for a manifold acted on by the identity group.
\end{example}

\begin{example} For the groupoid $G\ltimes M$ this gives the usual definition of an equivariant vector bundle.
 The \emph{tangent bundle} $T\Xx$ of an orbifold $\Xx$ is a
   orbibundle over $\Xx$.
\end{example}

\begin{example} If $U=[V/G]$ is a local chart (namely the restriction of the groupoid to a very small neighborhood), then a
   corresponding local uniformizing system for $T\Xx$ will be $[TV/G]$ with
   the action $g\cdot(x,v)=(gx,dg_x(v))$.
\end{example}

\begin{definition} Given an orbifold $\Xx$ we say that the space $X=\Xx_1 / \sim$ is its coarse topological space, or quotient space. Here $x\sim y$ whenever there is an arrow from $x$ to $y$. We will often write $\pi\colon \Xx_0 \to X$ to denote the canonical projection. \end{definition}

\begin{definition} Given a point $x\in X$ and an open neighborhood $x\in U \subseteq X$ we define $\Xx_U$ to be the restricted groupoid, namely its objects are $V=\pi^{-1} U$ and its arrows are all arrows $\alpha$ such that both $\pi(s(\alpha))$ and $\pi(t(\alpha))$ are in $U$. It is easy to show that for a sufficiently small $U$ we have that $\Xx_U$ is isomorphic to $[V/G]$ for some finite group $G$ acting on the manifold $V$. Such orbifold $[V/G]$ is called a local orbifold chart, or sometimes, a uniformizing system. An orbifold is called effective if at every point of $X$ we can find a local orbifold chart where the action of $G$ in $V$ is effective. \end{definition}

    Similarly the \emph{frame bundle} $P(\Xx)$ is a principal orbibundle over
   $\Xx$. The local uniformizing system is $U\times O(n) /G$ with local
   action $g\cdot(x,A)=(gx,dg\circ A)$.
 Notice that if the orbifold is effective then $P(\Xx)$ is
   always a \emph{smooth manifold} for the local action is free
   and $(s,t)\colon\Xx_1 \to\Xx_0 \times\Xx_0$ is one-to-one.
    From this we deduce that $\Xx = [P(\Xx)/O(n)]$. This proves the following proposition.

    \begin{proposition} Every effective orbifold arises from the almost free action of a Lie group on a manifold. \end{proposition}

\section{The Algebraic Topology of Orbifolds.}

 Define $$\Xx^{(n)} := \underbrace{\Xx_1 \timests \cdots \timests
\Xx_1}_{n}.$$ In the case in
which $\Xx_1$ is a set then $\Xx^{(n)}$ is the set of sequences
$(\gamma_1, \gamma_2, \ldots, \gamma_n)$ so that we can form the
composition $\gamma_1 \circ \gamma_2 \circ \cdots \circ \gamma_n$.

With  this data we can form a simplicial set \cite{Segal1}.

\begin{definition} A \emph{(semi-)simplicial set (resp. group, space, scheme)}
$X_\bullet$ is a sequence of sets $\{X_n\}_{n\in \naturals}$
(resp. groups, spaces, schemes) together with maps
$$
    X_0 \leftrightarrows X_1 \leftrightarrows X_2
    \leftrightarrows \cdots \leftrightarrows X_m \leftrightarrows
    \cdots
$$
$$
   \partial_i\colon X_m \to X_{m-1}, \ \ \ \ s_j\colon X_m\to
   X_{m+1}, \ \ \ \ 0\leq i,j \leq m.
$$
called \emph{boundary} and \emph{degeneracy} maps, satisfying
\begin{eqnarray*}
\partial_i \partial_j &=& \partial_{j-1} \partial_{i} \ \ \ \mbox{if $i<j$} \\
 s_i s_j &=& s_{j+1} s_i \ \ \ \mbox{if $i<j$} \\
\partial_i s_j &=& \left\{ \begin{array}{ll}
                            s_{j-1} \partial_i & \mbox{if $i<j$}\\
                            1 & \mbox{if $i=j,j+1$}\\
                            s_j \partial_{i-1} & \mbox{if
                            $i>j+1$}\\
                           \end{array}
                   \right.
\end{eqnarray*}

The nerve of a category (following Segal \cite{Segal1}) is a semi-simplicial set $N\CC$ where the
objects of $\CC$ are the vertices, the morphisms the 1-simplices, the triangular
commutative diagrams the 2-simplices, and so on.
\end{definition}

 For a category coming from a
groupoid then the corresponding simplicial object will satisfy
$N\CC_n=X_n=\Xx^{(n)}$.

We can define the boundary maps $\partial_i : \Xx^{(n)} \to
\Xx^{(n-1)}$ by:
\begin{displaymath}
\partial_i(\gamma_1, \dots , \gamma_n) = \left\{
 \begin{array}{ll}
(\gamma_2, \dots , \gamma_n) & \mbox{if $ i=0$} \\
(\gamma_1, \dots, m(\gamma_i,\gamma_{i+1}), \dots , \gamma_n) & \mbox{ if $ 1 \leq i \leq n-1$} \\
(\gamma_1, \dots, \gamma_{n-1}) & \mbox{if $ i =n $} \\
\end{array}
\right.
\end{displaymath}
and the degeneracy maps by
$$   s_j(\gamma_1,\ldots,\gamma_n)=\left\{ \begin{array}{ll}
                                          (e(s(\gamma_1)),\gamma_1,\ldots,\gamma_n)&\mbox{for
                                          $j=0$} \\
                                          (\gamma_1,\ldots,\gamma_j,e(t(\gamma_j)),
                                          \gamma_{j+1},\ldots,\gamma_n)&\mbox{for
                                          $j\geq1$}
                                         \end{array}
                                  \right.
$$

 We will write $\Delta^n$ to denote the standard $n$-simplex in
$\real^n$. Let $\delta_i\colon\Delta^{n-1}\to\Delta^n$ be the
linear embedding of $\Delta^{n-1}$ into $\Delta^n$ as the $i$-th
face, and let $\sigma_j\colon\Delta^{n+1}\to\Delta^{n}$ be the
linear projection of $\Delta^{n+1}$ onto its $j$-th face.

\begin{definition}The \emph{geometric realization} $|X_\bullet|$
of the simplicial object $X_\bullet$ is the space
$$ |X_\bullet|=\left.\left(\coprod_{n\in\naturals} \Delta^n \times
X_n \right)\right/ \begin{array}{c}
                    (z,\partial_i(x))\sim(\delta_i(z),x)\\
                    (z,s_j(x))\sim(\sigma_j(z),x)
                   \end{array}
$$
Notice that the topologies of $X_n$ are relevant to this
definition.
\end{definition}

 The simplicial object $N \CC$ determines $\CC$ and its
topological realization is called $B \CC$, the \emph{classifying
space of the category}. Again in our case $\CC$ is a
\emph{topological category} in Segal's sense.

\begin{definition} For a groupoid $\Xx$ we will call
$B\Xx=|N\Xx|$ the \emph{classifying space of the orbifold}.
\end{definition}

The following proposition establishes that $B$ is a functor from
the category of groupoids to that of topological spaces. Recall
that we say that two morphisms of groupoids are Morita related if
the corresponding functors for the associated categories are
connected by a morphism of functors.

\begin{proposition}   A morphism of groupoids $\Xx_1 \to \Xx_2$ induces a continuous
   map $B\Xx_1 \to B\Xx_2$. Two morphism that are Morita
   related will produce homotopic maps. In particular a Morita
   equivalence $\Xx_1 \sim \Xx_2$ will induce a homotopy
   equivalence $B\Xx_1\simeq B\Xx_2$. This assignment is
   functorial.
\end{proposition}

\begin{example} For the groupoid $\bar{G}=(\star \times G
\rightrightarrows \star)$ the space $B\bar{G}$ coincides with the
classifying space $BG$ of $G$.

Consider now the groupoid $\Xx=(G\times G \rightrightarrows G)$ where $s(g_1,
g_2)=g_1$, $t(g_1,g_2)=g_2$ and $m((g_1,g_2);(g_2,g_3))=(g_1,g_3)$ then it is easy to
see that $B\Xx$ is contractible and has a $G$ action. Usually $B\Xx$ is written $EG$

A morphism of groupoids $\Xx \to \bar{G}$ is the same thing as a principal $G$ bundle
over $\Xx$ and therefore can be written by means of a map $G \times G \to G$. If we
choose $(g_2,g_2)\mapsto g_1^{-1} g_2$ the induced map of classifying spaces
$$ EG \longrightarrow BG $$
is the universal principal $G$-bundle fibration over $BG$.

\end{example}

\begin{example} Consider a smooth manifold $X$ and a good open cover $\UU=\{U_{\alpha}\}_\alpha$.
Consider the groupoid $\GG=(\GG \rightrightarrows \GG_0)$ where $\GG_1$ consists on the
disjoint union of the double intersections $U_{\alpha\beta}$. Segal calls $X_U$ the corresponding topological category. Then he proves that $B\GG = BX_U \simeq X$.

 If we are given a principal $G$ bundle over $\GG$ then we have a morphism
$\GG\to\bar{G}$ of groupoids, that in turn induces a map $X\to BG$. Suppose that in
the previous example we take $G=\GLC{n}$. Then we get a map $X\to B\GLC{n}=BU$.
\end{example}

\begin{example}   Consider a groupoid $\Xx$ of the form $M\times G \rightrightarrows M$ where $G$ is
   acting on $M$ continuously. Then $B\Xx \simeq EG\times_G M$ is the Borel
   construction for the action $M \times G \to M$.
\end{example}

\begin{definition} The fundamental group of $\Xx$ is defined to be $\pi_1(\Xx) = \pi_1 (B\Xx)$. Similarly for the cohomology $H^*(\Xx)=H^*(B\Xx)$.\end{definition}

This last definition of cohomology is a bit too naive whenever we have obtained our orbifold by some geometric procedures. For example, as the space of solutions of algebraic equations. We will return to this issue later once we have the perspective given to us by topological quantum field theories.

\section{Loop Orbifolds}

The loop space is slightly more complicated in the case of an orbifold.

To generalize this situation to an orbifold $\Xx$ (replacing the
r\^ole of $M$ above), we must be able to say what is the candidate
to replace $\LL M$. This was done for a general orbifold in
\cite{LupercioUribeLoopGroupoid}. The basic idea is that to a groupoid $\Xx$ we must assign a new
(infinite-dimensional) groupoid $\Loop \Xx$ that takes the place
of the free loopspace of $M$ in a functorial manner $$\Loop \colon \Orbifolds \to \Orbifolds.$$

In  \cite{OrbifoldStringTopology} we prove the following.

\begin{theorem} The functor $\Loop$ defined in
\cite{LupercioUribeLoopGroupoid} commutes with the
functor $B$ from groupoids to spaces defined in the previous section.
Namely there is an homotopy equivalence $$ B\Loop \Xx \simeq \LL B
\Xx.$$
\end{theorem}

In the case in which $\Xx=[M/G]$, we proved that $\Loop \Xx$
admits a much smaller and very concrete model defined as follows.
The objects of the loop groupoid are given by $$(\Loop \Xx)_0 :=
\bigsqcup_{g \in G} \PPP_g,$$ where $\PPP_g$ is the set of all
pairs $(\gamma,g)$ with  $\gamma :\real \to X$ and $g \in G$ with
$\gamma(t)g = \gamma(2\pi + t)$. The space of arrows of the loop groupoid is $$(\Loop \Xx)_1 :=
\bigsqcup_{g \in G} \PPP_g  \times G,$$ and the action of $G$ in
$\PPP_g$ is by translation in the first coordinate and conjugation
in the second; that is, a typical arrow in the loop groupoid looks
like
$$(\gamma,g) \stackrel{((\gamma,g);h)}{\longrightarrow} (\gamma \cdot h, h^{-1} gh),$$
or pictorially:

\begin{center}
\includegraphics[width= 3.5 in, height= 1.3 in]{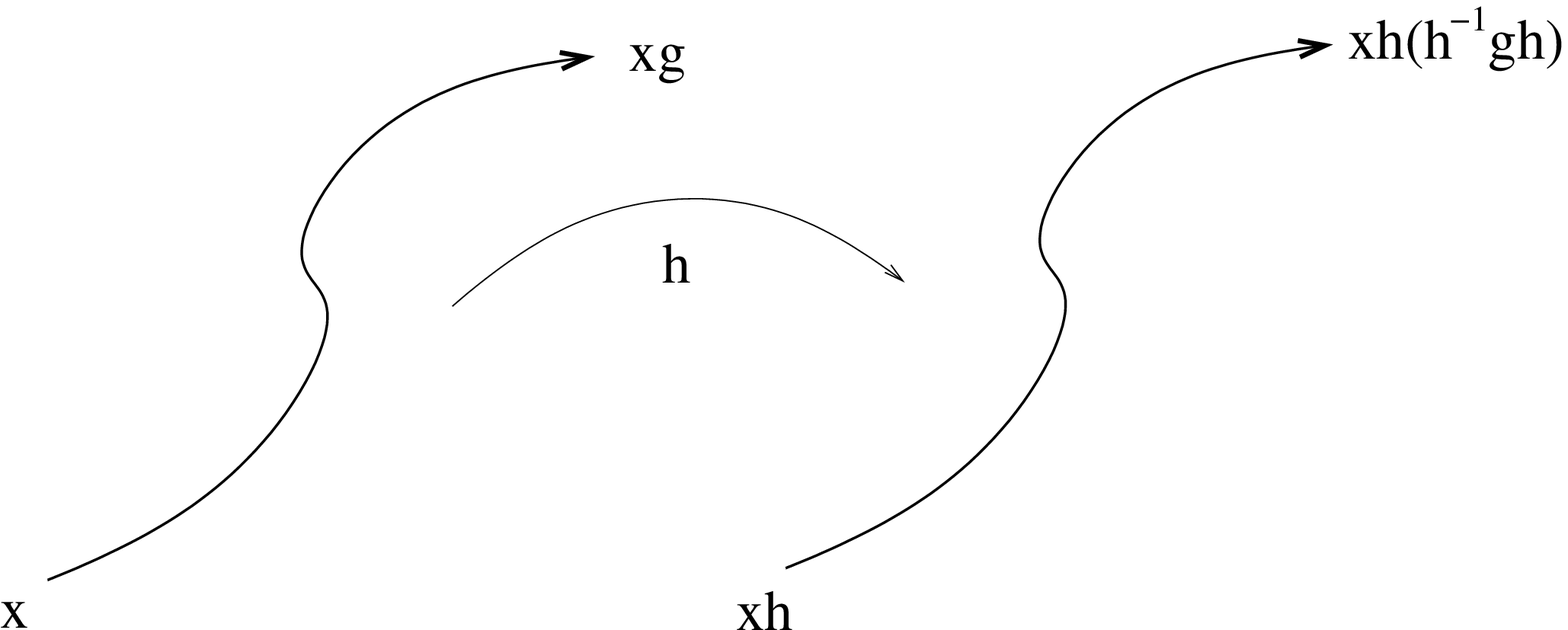}
\end{center}
\begin{theorem}[The Localization Principle \cite{dFLNU}]
Let $\Xx$ be an orbifold and $\Loop \Xx$ its loop orbifold. Then the
fixed orbifold under the natural circle action by rotation of loops is
\begin{equation}\label{LocalizationPrinciple}
(\Loop\Xx)^{S^1} = I(\Xx)
\end{equation}
where the groupoid $I(\Xx)$ has as its space of objects
$$ I(\Xx)_0 = \{ \alpha \in \Xx_1 \colon s(\alpha) = t(\alpha) \} = \coprod_{m \in \Xx_0} \Aut_\Xx (m)$$ and its space of arrows is $$I(\Xx)_1 = Z(I(\Xx_0))=\{ g\in \Xx_1 \colon \alpha \in I(\Xx)_0 \Rightarrow g^{-1} \alpha g \in I(\Xx)_0 \},$$ a typical arrow in $I(\Xx)$ from
$\alpha_0$ to $\alpha_1$ looks like
$$
        \xymatrix{
        \circ \ar@(ul,dl)[]|{\alpha_0} \ar@/^/[rr]|{g}
        &&\circ \ar@(dr,ur)[]|{\alpha_1^{-1}} \ar@/^/[ll]|{g^{-1}}
}$$
\end{theorem}

While for a smooth manifold the space of constant maps is $$M=(\LL M )^{S^1}$$ we have in
contrast $$\Xx \subset I(\Xx) = (\Loop\Xx)^{S^1}.$$

In \cite{LupercioUribeGhost} we define the \emph{ghost loop
space} $\LL_s B\Xx$ as the subspace of elements $\gamma \in \LL B
\Xx$ so that the composition with the canonical projection
$\pi_\Xx \colon B\Xx \to X$, $\pi_\Xx \circ \gamma$ is constant.
In that paper it is proved the homotopy equivalence
$$B I(\Xx) \simeq \LL_s B \Xx.$$

\section{Orbifold TQFTs}

\subsection{Finite Groups.} \label{finitegroups}
    There is a very beautiful example of a topological quantum field theory (TQFT) due to Dijkgraaf and Witten \cite{DijkgraafWitten, SegalStanford, CohenVoronov}. This is a $(n+1)$-dimensional TQFT $(H^G,\Psi^G,Z^G)_{n+1}$ associated to a finite group $G$. In this model we have
\begin{itemize}
    \item $\FF(Y) = [Y,BG] = \Bun_G(Y)$, where $\Bun_G(Y)$ is the set isomorphism classes of $G$-principal bundles on $Y$.
    \item $H^G(Y)=\Map(\Bun_G(Y), \complex)$. Here we remark that $\Bun_G(Y) \cong \Hom_\integer(\pi_1(Y),G)/\sim$, this last bijection being induced by the holonomy of the bundle. The symbol $\sim$ denotes conjugation.
    \item For a boundaryless $Y$ we have $Z^G(Y)=|\Hom(\pi_1(Y);G)|/|G|$.
    \item If $\partial Y=Z$ has no output boundary then for each $P\in\Bun_G(Z)$ we have $$\Psi_Y(P) = \sum_{Q\in\Bun_G(Y),\ Q|Z=P} \frac{1}{|\Aut(Q)|} \in \complex $$
\end{itemize}
Segal has shown that when de dimension of the model is $1+1$ then we have
\begin{itemize}
    \item The Frobenius algebra $(A_G,\theta_G)$ associated to $(H^G,\Psi^G,Z^G)_{1+1}$ is isomorphic to the center of the group algebra $\complex[G]$, with trace $$\theta_G\left(\sum_g \lambda_g g\right) = \frac{1}{|G|}\lambda_1.$$
    \item For a boundaryless Riemann surface $\Sigma$ we have $$Z(\Sigma)= |G|^{2g-2} \sum_V \frac{1}{(\dim V)^{2g-2}}$$ where $g$ is the genus of $\Sigma$ and $V$ runs through irreducible representations of $G$.
\end{itemize}

\subsection{Orbifold String Topology}
Orbifold string topology \cite{OrbifoldStringTopology} is a quantum field theory that generalizes simultaneously the models of Dijkgraaf-Witten and of Chas-Sullivan described above, actually the theory interpolates between those two theories.

We fix a compact oriented orbifold $\Xx$. This is a (positive boundary) (1+1)-dimensional topological quantum field theory with fields $$\FF(Y) = \Map(Y,\Xx)$$ the moduli space of all orbifold morphisms from $Y$ to $\Xx$. We have replaced the manifold $M$ of string topology with an orbifold $\Xx$. The Dijkgraaf-Witten theory is recovered when we consider the orbifold $\Xx=[\bullet/G]$.

    In \cite{OrbifoldStringTopology} we have introduced the string topology of $\Loop \Xx$. In order to do this we can consider (stack) orbifold correspondences. Such an orbifold correspondence is a diagram in the category of orbifolds of the form
$$\xymatrix{
\ & \Zz \ar[dr] \ar[dl] & \  \cr
\Xx  & \ & \Yy \cr
}
$$ We need of course the concept of smooth degree $-d$ correspondence in the category of infinite dimensional $C^\infty$ orbifolds to mimic the arguments of the previous subsections. In any case whenever $\Xx$ is a $d$ dimensional oriented orbifold, the output of this procedure is a degree $-d$ product on the homology of $\Loop \Xx$, $$ H_*(\Loop \Xx) \otimes H_*(\Loop \Xx) \to H_{*-d}(\Loop \Xx).$$

    In \cite{OrbifoldStringTopology} we consider mostly global orbifolds, namely orbifolds of the form $\Xx=[M/G]$ where $G$ acts in an orientation preserving fashion on the compact, closed smooth manifold $M$. In this case we have that $A_{\Loop \Xx} = H_*(\Loop \Xx) \cong H_*(\LL(M\times_G EG))$. Here the main idea comes from string theory, and it is to consider the elements of $G$ as $0$-branes, in the form of boundary conditions for the strings. In any case, using classical algebraic topology we prove the following.

\begin{theorem} Let $\Xx=[M/G]$, then $A_{\Loop \Xx} \cong H_*(\LL(M\times_G EG); \rational)$ has the structure of a Batalin-Vilkovisky algebra, and moreover
\begin{itemize}
    \item When $G=\{1\}$ and for arbitrary $M$ then $A_{\Loop \Xx}$ coincides with the Chas-Sullivan BV-algebra.
    \item More interestingly, when $M=\{m_0\}$ is a single point and for arbitrary $G$ then $A_{\Loop \Xx}$ coincides with the Dijkgraaf-Witten Frobenius-algebra.
\end{itemize}
\end{theorem}

\subsection{Chen-Ruan Orbifold Stringy Cohomology}
The Chen-Ruan theory \cite{ChenRuan} is a (1+1)-dimensional TQFT whose fields are of the form
$$\FF(Y)=\Hol^\odot(Y,\Xx).$$ The constant holomorphic orbifold maps from $Y$ to $\Xx$. Recall that the localization principle states that for the boundary values we have $$\FF(S^1)= \Map^\odot(S^1,\Xx) = I(\Xx),$$ is the inertia orbifold.
The correspondence associated to the pair of pants $P$ is
$$\xymatrix{
\ & \FF(P) \ar[dr] \ar[dl] & \  \cr
I(\Xx)\times I(\Xx)  & \ & I(\Xx) \cr
}
$$
and it gives a product $$H^*(I(\Xx))\otimes H^*(I(\Xx)) \to H^*(I(\Xx)).$$
We should mention that $\FF(P)$ has a virtual fundamental class of \emph{rational} degree (called the \emph{degree shifting} in the terminology of Chen and Ruan). Considering this product, and the degree shift the Chen-Ruan cohomology of an orbifold is defined as $$H^*_{\mathrm{Chen-Ruan}}(\Xx) :=  H^*(I(\Xx)).$$
For a survey of its remarkable properties and conjectures we recommend the reader to consult \cite{RuanStringy1, RuanStringy2}.

\section{Acknowledgments} This paper is based on the lectures given by the first author in Villa de Leyva, Colombia in August 2005. We would like to thank the organizers of the summer school for the invitation to deliver the lectures.

The authors would like to thank Alejandro Adem, Ralph Cohen, Dan Freed, Graeme Segal and Constantin Teleman for conversations that have clarified diverse aspects of the subjects covered here.

Both authors woud like to thank the MSRI for its hospitality during the preparation of this manuscript, the second author would also like to thank the Max Planck Institut in Bonn.

\bibliographystyle{amsplain}
\bibliography{OrbTQFTbib}

\end{document}